\documentclass[fleqn,twoside]{article}

\usepackage[headings]{espcrc2}
\usepackage{graphicx}
\usepackage[figuresright]{rotating}
\usepackage{amsmath}
\usepackage{epsfig}

\newcommand{\be}{\begin{equation}}
\newcommand{\ee}{\end{equation}}
\newcommand{\bea}{\begin{eqnarray}}
\newcommand{\eea}{\end{eqnarray}}
\newcommand{\eps}{\varepsilon}
\newcommand{\ep}{\varepsilon}

\def \litwo   {{\rm{Li_2}}}
\def \litr    {{\rm{Li_3}}}
\def \litt {{\rm{S_{2,2}}}}
\def \liot {{\rm{S_{1,2}}}}
\def \lifo {{\rm{Li_4}}}

\title{%
{\small
\tt DESY 06-154
\\[-2mm]
SFB/CPP-06-40}
\\[3mm]
Planar two-loop master integrals for massive Bhabha scattering: 
\\
$N_f=1$ and $N_f=2$}

\author{
Stefano~Actis%
\address[ZE]{Deutsches Elektronen-Synchrotron, DESY, Platanenallee 6, D-15738 Zeuthen, Germany}, 
Micha{\l}~Czakon%
\address{Institut f\"ur Theoretische Physik und Astrophysik, Universit\"at W\"urzburg,
         Am Hubland,\\ D-97074 W\"urzburg, Germany}%
\address[KA]{Institute of Physics, University of Silesia, Uniwersytecka 4, PL-40007 Katowice, Poland},
Janusz~Gluza%
\addressmark[KA],
Tord~Riemann%
\addressmark[ZE]
}

\runtitle{}
\runauthor{}

\begin{document}

\begin{abstract}
Recent developments in the computation of two-loop master integrals for
massive Bhabha scattering are briefly reviewed.
We apply a method based on expansions of exact Mellin-Barnes representations and
evaluate all planar four-point master integrals in the approximation of small electron mass at fixed scattering angle for the one-flavor case.
The same technique is employed to derive and evaluate also all two-loop
masters generated by additional fermion flavors.
The approximation is sufficient for the determination of QED two-loop corrections for Bhabha scattering in the kinematics planned to be used for the luminosity determination at the ILC.
\end{abstract}
\maketitle
\pagestyle{empty}
\thispagestyle{empty}

\allowdisplaybreaks

\section{Introduction\label{section-intro}}
Bhabha scattering is the process employed to 
measure the luminosity at electron-positron colliders,
because of its clear experimental signature.
At machines operating at a $1 - 10$ $\rm{GeV}$ centre-of-mass energy $\sqrt{s}$, 
the relevant kinematic region is that of large-angle Bhabha scattering.
Small-angle Bhabha scattering, instead,  is an invaluable
luminosity monitor at high-energy colliders in the TeV region.

In order to minimize the luminosity error,
a precise theoretical computation of radiative
corrections to the Bhabha-scattering cross section is required.
The electroweak next-to-leading order (NLO) corrections
to Bhabha scattering were computed
a long time ago in \cite{Consoli:1979xw}.
In recent years, studies
have been focusing on pure quantum electrodynamics (QED) contributions
beyond the one-loop level.
The two-loop virtual corrections for massless
electrons were obtained in \cite{Bern:2000ie}.
However, this result was not immediately useful since the available Monte Carlo programs
employ a non-vanishing electron mass $m$.

The virtual and real second-order contributions to
Bhabha scattering, enhanced by factors of $\ln (s / m^2)$
and $\ln^2 (s / m^2)$ were completed in \cite{Arbuzov:1995vi,Arbuzov:1995vj,Arbuzov:1998du,Glover:2001ev}.
This result was recently improved in \cite{Penin:2005kf,Penin:2005eh,Penin:2005iv},
where the photonic  non-logarithmic term
was evaluated at leading order in the ratio $m^2 \slash s$.
The diagrams with fermion loops remained uncovered in this approach.

An important breakthrough in the field was the
use of the Laporta-Remiddi algorithm (\cite{Laporta:1996mq,Laporta:2001dd}),
in order to reduce the Bhabha-scattering
cross section to a few Master Integrals
(MIs).
The technique of differential equations
proved useful in evaluating several MIs
(see i.e. \cite{Bonciani:2003cj,Czakon:2004wm,Czakon:2005gi}).
The results were represented 
in terms of Harmonic Polylogarithms (HPLs) introduced in \cite{Remiddi:1999ew} or
of Generalized Harmonic Polylogarithms (GPLs)
(details in the context of Bhabha scattering can be found in
\cite{Czakon:2005jd} and in references therein).
These results led eventually to the exact result
of \cite{Bonciani:2004gi,Bonciani:2004qt} for the virtual and real next-to-next-to-leading
order (NNLO) corrections
to the Bhabha-scattering cross section
involving one electron loop.
Non-approximated expressions for all NNLO contributions,
except for double box diagrams, can be found in
\cite{Bonciani:2005im,Bonciani:2006qu}.
The MIs for the loop-by-loop contributions were studied e.g. in \cite{Fleischer:2006ht}. 

\begin{table*}[htb]
\label{tab-boxes}
\caption{
The $N_f=1$ four-point master integrals entering the six basic two-loop box diagrams.
NP denotes non-planar topologies, and references with a dagger give divergent parts only.}
\vspace{2mm}
\begin{tabular}{|l|c|c|c|c|c|c|l|}
\hline
MI &                B1 & B2 & B3 & B4 & B5 & B6& \\
\hline \hline
{\tt B7l4m1} &            +  &  -- & --  & --  &  -- & -- & \cite{Smirnov:2001cm,Czakon:2006pa} \\
{\tt B7l4m1N} &           +  &  -- & --  & --  &  -- & --
&\cite{Heinrich:2004iq,Czakon:2006pa}
\\ 
{\tt B7l4m2} &            --
&  + & --  & --  &  -- & --  & 
\cite{Heinrich:2004iq,Czakon:2006pa}
\\ 
{\tt B7l4m2[d1-d3]} &     --  & +  & --  & --  &  -- & --
&\cite{Czakon:2006pa}
\\ 
{\tt B7l4m3}  &            --  & --  &
+  & --  &  -- & --  & 
NP \cite{Heinrich:2004iq}$^{\dagger}$
\\
{\tt B7l4m3[d1-d2]} &     --  & --  & +  & --  &  -- & --  &
NP
\\

\hline
{\tt B6l3m1} &            +  & --  & +  & --  &  -- & --  &\cite{Czakon:2006pa}\\
{\tt B6l3m1d} &           +  & --  & +  & --  &  -- & --  &\cite{Czakon:2006pa}\\
{\tt B6l3m2} &            --  & +  & --  & +   &  -- & --  &\cite{Czakon:2006pa}\\
{\tt B6l3m2d} &           --  & +  & --  & +   &  -- & --  &\cite{Czakon:2006pa}\\
{\tt B6l3m3} &            --  & --  & +  & --  &  -- & --  &
NP
\\
{\tt B6l3m3[d1-d5]} &     --  & --  & +  & --  &  -- & --  &
NP
\\
\hline
{\tt B5l2m1} &            +  & --  & +  & --  &  -- & -- & \cite{Czakon:2004tg} \\
{\tt B5l2m2}  &           --  & +  & --  & +  &  -- & +
&\cite{Czakon:2006pa}\\ 
{\tt B5l2m2[d1-d2]} &     --  & +  & --  & +  &  -- & +
&\cite{Czakon:2006pa}\\ 
{\tt B5l2m3}    &         +  & --  & +  & --  &  -- & --  &\cite{Czakon:2006pa}\\
{\tt B5l2m3[d1-d3]} &     +  & --  & +  & --  &  -- &
--&\cite{Czakon:2006pa}\\ 
{\tt B5l3m}     &        --  & +  &  +  & +  & --  & --  &\cite{Czakon:2006pa}\\
{\tt B5l3m[d1-d3]}&      -- & +  &  +  & +  & --  & --  &\cite{Czakon:2006pa}\\
{\tt B5l4m}    &          --  & +  & +  & +  &  + & --  & \cite{Bonciani:2003cj} \\
{\tt B5l4md} &            --  & +  & +  & +  &  + & --  & \cite{Czakon:2004tg}
\\
\hline
\end{tabular}
\end{table*}

\begin{figure}
\label{fig-nf2}
\includegraphics[height=9cm,width=9.cm,angle=0,scale=0.8]{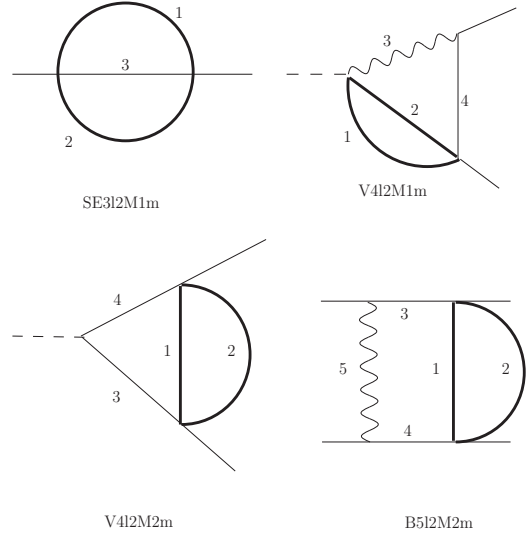}
\vspace*{-0.3cm}
\caption{The topologies of the eight master integrals
for the heavy-fermion corrections.
Bold lines represent heavy fermions.}
\end{figure}

The complete set of the needed master integrals is known from
\cite{Czakon:2004wm}.
Table \ref{tab-boxes} reproduces all two-loop box master integrals for $N_f = 1$.
Notations are exactly those of \cite{Czakon:2004wm}.
The {\tt B7l4m3d2} is, e.g., a box MI with 7 internal lines (7l), four of them being massive (4m), with a higher power of one of the numerators (a line being dotted (d));
it is one of several such topologies and of several ones with dots, so '3d2'.
In order to improve the Bhabha-scattering theoretical prediction,
we investigate two classes of NNLO QED corrections.
In Section \ref{section-planar},
we briefly discuss a method based on expansion of Mellin-Barnes (MB)
representations (\cite{Usyukina:1975yg,Boos:1991rg,Smirnov:1999gc,Tausk:1999vh})
and review the results of \cite{Czakon:2006pa},
where all planar two-loop box MIs were obtained.
The non-planar MIs  are indicated in Table \ref{tab-boxes}.
In Section \ref{section-nf2}, we  apply the same method
to evaluate the MIs arising from diagrams containing 
heavy fermions, like muons and tau-leptons;
in the following, we will call them the $N_f > 1$ contributions.
Their topologies are shown in Figure \ref{fig-nf2}.

The MB-representations are valid for arbitrary kinematics.
Although their actual evaluations are restricted to the high-energy limit (small lepton masses at fixed scattering angle), 
they are well suited for practical applications.
When dealing with $N_f > 1$ MIs,
a second fermion mass $M$ is involved.  
Since our
purpose is to evaluate the complete QED Bhabha-scattering cross section at high
energies, we assume a hierarchy of all three
scales, namely $m \ll M \ll s,t$, where $t$ is the usual
Mandelstam invariant related to the scattering angle. 
With the summation techniques described in Section \ref{section-techn}, divergent parts have been evaluated exactly.
Note that the
treatment of hadronic contributions is a separate problem, which is
better solved by using dispersion relations
(see \cite{Kniehl:1988id}). 

\vfill

\begin{figure}[htbp]
\label{fig-diagnf2}
\vspace*{-0.1cm}
\hspace*{0.5cm}
\includegraphics[height=10.5cm,width=8.cm,angle=0,scale=0.8]{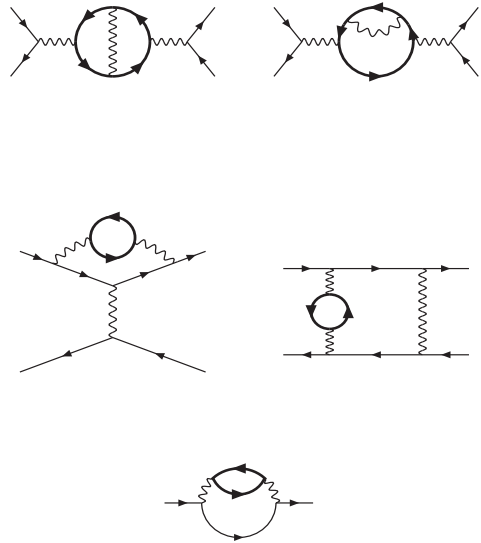}
\vspace*{-0.9cm}
\caption{The diagrams for Bhabha scattering with two fermion flavors.
Internal fermionic loops represent heavy leptons and the other fermion lines are electrons.}
\end{figure}

\section{The planar two-loop boxes for $N_f=1$\label{section-planar}}
There are 24 planar two-loop box MIs to be determined for the $N_f=1$ case, see Table \ref{tab-boxes}.
So far, we could not determine all of them analytically with the exact mass dependence; the status is reviewed in \cite{Czakon:2006hb}.
For this reason, we decided to treat these MIs uniquely in the approximation of small $m$ at fixed scattering angle.
The results have been published in the mean time \cite{Czakon:2006pa}, so we will make here only few introductory remarks and show one example.
In the list of MIs, we preferred to include Feynman integrals without any numerators.
The pragmatic reason was the independence of their defintion on the internal flow of momenta.
It is sufficient to indicate the lines with dots.
Performing explicit calculations, one is, of course, faced with the observation that the singularities of dotted MIs and those with numerators are quite different.
Evaluating the small mass expansions, we preferred in some cases to treat instead of dotted integrals those with numerators.
Due to unique algebraic relations between all the integrals, there is no principal difference in these two approaches, and further details are discussed in   \cite{Czakon:2006pa}.
Another observation concerns the MB-representations.
In principle, one may use the representations for the basic 7-line boxes as given e.g. in \cite{Smirnov:book4} and shrink lines.
However, when calculating MIs with numerators, additional representations have to be determined.
We observed further, that it is sometimes not evident how to get an effective represenation for dotted MIs from more general ones.
For the MI {\tt B5l2m2} (a 5-liner) we got, by shrinking of lines in {\tt B7l4m1} (first planar double box)and after expanding in $\eps$, 11 MB-integrals with at most 4 integrations.
From our direct derivation, we got 4 integrals, at most 3-dimensional.
For the related dotted MI {\tt B5l2m2d2}, we got from line shrinking 102 integrals, and by direct derivation only one, again 3-dimensional.

\begin{figure}
  \begin{center}
    \epsfig{file=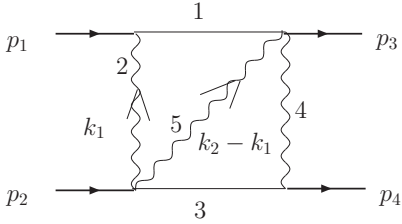,width=6cm}
  \end{center}
  \caption{\label{5lin}
The 5-line topology {\tt B5l2m3}. The momentum distribution has been chosen to
make the derivation of the MB representation easier.
}
\end{figure}

As an example, we reproduce here for the MI B5l2m3($k_2 p_3$), shown in Figure \ref{5lin}, the basic $d$-dimensional MB-represenation:
\begin{eqnarray}
{\tt B5l2m3(p_e \cdot k_2)} 
\nonumber \\ 
=
\frac{ (-1)^{a_{12345}}\;
e^{2\ep\gamma_E}}{ \prod_{j=1}^5\Gamma[a_{i}]
   \Gamma[4  - 2\ep- a_{123} ](2\pi i)^4}
\nonumber\\
\int_{-i \infty}^{+i \infty} d \alpha \int_{-i \infty}^{+i \infty}
d \beta \int_{-i \infty}^{+i \infty} d \gamma \int_{-i \infty}^{+i \infty} d \delta
\nonumber \\ 
(-s)^{\gamma} \;(-t)^{(4 - 2 \epsilon - a_{12345} - \beta - \delta  - \gamma)} \nonumber
\\
\Gamma[-\beta] \;\Gamma[-\gamma] \;\Gamma[-\delta] \; \Gamma[a_3 + \alpha + 2 \;\beta] \; \nonumber\\
\frac{\Gamma[2 - \epsilon - a_{45} + \alpha - \delta  - \gamma]}
{\Gamma[7 - 3 \epsilon- a_{12345} - \beta]  } \nonumber \\
\frac{\Gamma[2 - \epsilon - a_{13} - \beta ]
\;\Gamma[2- \epsilon - a_{23} - \alpha - \beta ]}
{ \Gamma[a_5 - \alpha + 2 \;\gamma] \;\Gamma[1 + a_5 - \alpha + 2 \;\gamma]} \nonumber \\
  \Gamma[-4 + 2 \;\epsilon + a_{12345} + \beta + \delta + \gamma] \nonumber \\
\biggl\{ \Gamma[4 - 2 \;\epsilon - a_{1235} - \beta - \delta  - \gamma] \nonumber \\
   \biggl[ (p_e \cdot p_2) \;\Gamma[1 + a_5 + \gamma] \;\Gamma[-\alpha + \gamma] 
\nonumber\\
- (p_e \cdot p_1) \;\Gamma[a_5 + \gamma] \;\Gamma[1 - \alpha + \gamma] \biggr]
\nonumber \\
\Gamma[a_5 - \alpha + 2 \;\gamma] \;\Gamma[1 + a_5 - \alpha + 2 \;\delta + 2 \;\gamma] \nonumber \\
+
   [(p_e \cdot p_3) - (p_e \cdot p_1)] \;
\nonumber \\
\Gamma[5- 2 \epsilon  - a_{1235} - \beta - \delta - \gamma] \nonumber \\
\Gamma[a_5 + \gamma] \;\Gamma[-\alpha + \gamma] \;\Gamma[1 + a_5 - \alpha + 2 \;\gamma] \;
\nonumber\\
    \Gamma[a_5 - \alpha + 2 \;(\delta + \gamma)] \biggr\}\nonumber
\label{numB5l2m3}
\end{eqnarray}

The small mass expansion of the result is:
\begin{eqnarray}
{\tt B5l2m3(k_2\cdot p_2)} 
 =\frac{1}{4}\left(\frac{s}{u}\right)^2 \biggl\{
{\rm L}^2\;(6\;x\;\zeta_2 
\nonumber \\
+ 2\;x\;\ln(x) + 2\;x^2\;\ln(x)
+ x\;\ln^2(x)) \nonumber \\
+{\rm L}\;(16\;x\;\zeta_2 - 8\;x^2\;\zeta_2
\nonumber \\
 - 4\;x\;\zeta_3 -
2\;\ln(x) + 2\;x^2\;\ln(x) \nonumber \\
+4\;x\;\zeta_2\;\ln(x) + 2\;x\;\ln^2(x) - 2\;x^2\;\ln^2(x)
\nonumber \\
 -
   12\;x\;\zeta_2\;\ln(1 + x) \nonumber \\
-2\;x\;\ln^2(x)\;\ln(1 + x)
- 4\;x\;\ln(x)\;{\litwo}( -x)
\nonumber \\
 + 4\;x\;{\litr}( -x)) \biggr\} \nonumber \\
+ \frac{1}{120} \left(\frac{s}{u}\right)^2 \biggl\{ +120\;\zeta_2
  \nonumber \\
+ 360\;x\;\zeta_2
\nonumber \\
- 120\;x^2\;\zeta_2 - 1560\;x\;\zeta_4 - 480\;x\;\zeta_3
\nonumber \\
-240\;x^2\;\zeta_3 - 240\;x\;\zeta_2\;\ln(x)
\nonumber \\
- 480\;x^2\;\zeta_2\;\ln(x) - 360\;x\;\zeta_3\;\ln(x)
\nonumber \\
+30\;\ln^2(x) + 60\;x\;\ln^2(x) - 30\;x^2\;\ln^2(x)
\nonumber \\
 -
 180\;x\;\zeta_2\;\ln^2(x) 
\nonumber \\
-20\;x\;\ln^3(x) - 20\;x^2\;\ln^3(x) - 5\;x\;\ln^4(x)
\nonumber \\
+ 720\;x^2\;\zeta_2\;\ln(1 + x)
\nonumber \\
 +120\;x\;\zeta_3\;\ln(1 + x)
\nonumber \\
 - 120\;x\;\zeta_2\;\ln(x)\;\ln(1 + x)
\nonumber \\
 +
120\;x^2\;\ln^2(x)\;\ln(1 + x) \nonumber \\
+
 180\;x\;\zeta_2\;\ln^2(1 + x) 
\nonumber \\
+ 30\;x\;\ln^2(x)\;\ln^2(1 + x)
 \nonumber \\
+
120\;x\;\ln(x)\;\liot 
(-x) \nonumber \\
+60\;x\;(-8\;\zeta_2 - \ln^2(x) + 2\;\ln(x)\;
\nonumber \\
(2\;x + \ln(1 + x)))\;{\litwo}( -x)
\nonumber \\
- 240\;x^2\;{\litr}( -x) +
 240\;x\;\ln(x)\;{\litr}( -x)
\nonumber \\
 - 120\;x\;\ln(1 + x)\;{\litr}( -x)
\nonumber \\
-360\;x\;{\lifo}( -x) - 120\;x\;{\litt}(-x)
\biggr\}\nonumber
\label{B5l2m2N1ms}
\end{eqnarray}
The expression is more complicated than those for the planar 7-line MIs concerning both the functions appearing as well as the dependence on all three Mandelstam variables $s,t,u$; the latter is typical for non-planar diagrams, where the {\tt B5l2m3} topology contributes.

\section{The master integrals for the $N_f > 1$ corrections\label{section-nf2}}
The differential Bhabha-scattering cross section with respect
to the solid angle $\Omega$ can be written
by means of an expansion in the fine-structure constant $\alpha$,
\be
\frac{d \sigma}{d \Omega}= \frac{d \sigma_0}{d \Omega} + 
\left(\frac{\alpha}{\pi}\right) \frac{d \sigma_1}{d \Omega} +
\left(\frac{\alpha}{\pi}\right)^2 \frac{d \sigma_2}{d \Omega}+\ldots,
\ee
where $\sigma_0$ is the Born contribution and $\sigma_i$ ($i=1,2,\ldots$)
represent the higher-order radiative corrections.
If we are interested in the NNLO virtual contributions,
we need to evaluate the diagrams of Figure 2, 
where the fermion self-energy is required for wave-function
renormalization.
Note that results for the photonic vacuum polarization diagrams
can be found in \cite{Kallen:1955fb}.

After interfering the two-loop amplitude with
the tree-level one, summing over the spins of the
final state and averaging over those of the initial
state, we get a large number of integrals.
We use the \texttt{DiaGen/IdSolver} \cite{Czakon:2004uu2}
implementation of the Laporta-Remiddi algorithm
\cite{Laporta:1996mq} in order to reduce all the needed
Feynman integrals to a limited set of MIs.
Apart from products of one-loop integrals,
we get the eight MIs of Figure \ref{fig-nf2}, as already pointed out  in \cite{Czakon:2004wm}.
The corresponding Feynman integrals with $n$ propagators $D_i$ are defined as follows:
\be
\label{feynmi}
D(\{\nu_i\}_n) = -
\frac{(e^\gamma)^{2\eps}}
{\pi^d}\int\frac{d^dk_1d^dk_2}{\prod_{i}^{n}D_i^{\nu_i}},
\ee
where $\gamma$ is the Euler-Mascheroni constant and we introduced
the shorthand notations
\begin{eqnarray}
\{\nu\}_n&\equiv& \nu_1,\ldots,\nu_n,\nonumber\\
\nu_{ab\ldots c}&\equiv&\nu_{a}+ \nu_{b} +\ldots + \nu_{c}.
\end{eqnarray}

In contrast to \cite{Czakon:2006pa}, we do not  consider MIs with 
scalar products in the numerators.
We have then to allow for higher powers of propagators.
Of course, there are algebraic relations between
MIs with scalar products in the numerators
and MIs with propagators raised to higher powers.

We construct our MB representations using the standard approach described in
\cite{Smirnov:book4}. The Feynman-parameter
integrals are derived one after the other for the two subloops. 
In each step we replace the sum over monomials in the Feynman parameters
by an appropriate MB representation. Due to the relatively simple structure of the
considered diagrams, it is easier to begin with the propagator-type subloop.

As far as the electron self-energy is concerned, we only need the sunrise
topology with the electron on its mass shell, $p_1^2=m^2$. 
A number of results for this mass configuration of the sunrise diagram can be found in the
literature.  
Analytic expressions  for the residues of the poles in dimensional regularization 
and the finite parts were given already
in \cite{Berends:1998vk}.
The explicit result  for  the ${\cal O} (\epsilon)$ terms, where $\epsilon \equiv (4-d)/2$
and $d$ are the space-time dimensions,
can be found in \cite{Argeri:2002wz}.
Note that the inclusion of the ${\cal O} (\epsilon)$ terms
for the sunrise MIs is mandatory when
deriving the complete squared amplitude,
since the reduction to MIs generates inverse powers of $\epsilon$.

For the self energy, we use our MB representation in order to reproduce the 
known result for the MIs.
Its general form, for arbitrary powers $\nu_i$ of
the propagators, is given by
\begin{eqnarray}\label{def:SUN}
&&\texttt{SE3l2M1m}(\{\nu\}_3) =
(m^2)^{4-\nu_{123}-2 \epsilon}
\nonumber \\
&&
\times \frac{(-1)^{\nu_{123}} e^{2 \gamma \epsilon}}{\prod_{i=1}^3 \Gamma({\nu_i})}
 \int \frac{dz}{2\pi i} \left( \frac{m^2}{M^2}\right)^{z+\nu_{12}-2+\epsilon}
\nonumber\\
&&\times\frac{\prod_{i=1}^6 \Gamma_i}
{\Gamma(2 z + \nu_{12}) \Gamma(z-\nu_3+4-2 \epsilon)}.
\end{eqnarray}
Furthermore, we defined
\begin{xalignat}{2}
\Gamma_1 &\equiv\Gamma(-z),      \nonumber\\
\Gamma_2&\equiv\Gamma(z+\nu_1),  \nonumber\\
\Gamma_3&\equiv\Gamma(z+\nu_2),   \nonumber\\ 
\Gamma_4&\equiv\Gamma(-z+\nu_3-2+\epsilon), \nonumber\\
\Gamma_5&\equiv\Gamma(2 z-\nu_3+4-2\epsilon), \nonumber\\ 
\Gamma_6&\equiv\Gamma(z+\nu_{12}-2+\epsilon).
\end{xalignat}
The integration contour is a straight line parallel
to the imaginary axis separating the poles
generated by $\Gamma_1$ and $\Gamma_4$ from
those coming from $\Gamma_2$, $\Gamma_3$, $\Gamma_5$ and $\Gamma_6$.

The two sunrise MIs are defined by the following values
for the powers of the propagators,
\begin{eqnarray}
\texttt{SE3l2M1m}&\equiv&\texttt{SE3l2M1m}(1,1,1),\nonumber\\
\texttt{SE3l2M1md}&\equiv&\texttt{SE3l2M1m}(1,2,1).
\end{eqnarray}

Having a MB representation at hand, one needs
to perform an analytic continuation in $\eps$
from a range where the integral is regular to the vicinity
of the origin, uncovering the singular structure on the way.
This is done
by an automatized procedure   implemented in the
Mathematica package {\tt MB.m} \cite{Czakon:2005rk}.
The resulting MB representations are verified
numerically in the Euclidean region against the sector
decomposition approach as described in \cite{Binoth:2003ak}.

For the sunrise MIs, a straightforward application of the Cauchy theorem
to the MB representation of Eq.~\eqref{def:SUN}
leads to a sum over residua which can be easily
evaluated.
Therefore, we reproduced the results of \cite{Argeri:2002wz}.
In general, however, one has to deal with multiple
MB representations.
For the vertex and box MIs, the evaluation of
the needed sums is far from being trivial.

As explained in the introduction, our purpose is
to calculate the integrals by assuming a hierarchy
of all scales, namely $m^2 \ll M^2 \ll s,t$.
First of all we 
identify the leading contributions in the electron mass
following the procedure described in \cite{Czakon:2006pa}.
Then, by using the Cauchy theorem to express
the integrals through sums over residua,
we evaluate these sums with the aid of {\tt XSUMMER} \cite{Moch:2005uc}.

For the sunrise MIs the results depend on one variable,
$R\equiv m^2 \slash M^2$,
and read as
\begin{eqnarray}
\texttt{SE3l2M1m}\ &=& \ M^2\ (m^2)^{-2 \epsilon} 
\nonumber\\
&&\times \Bigl[ \sum_{k=-2}^{1}\ S_k\ \epsilon^k\ +\ {\cal O} (\epsilon^2) \Bigr], 
\nonumber\\
S_{-2}&=&  1  , 
\\
 S_{-1}&=& 3 + 2 \ln \left(R\right) ,\nonumber\\
 S_0 &=&
7 + \zeta_2  + 6 \ln \left(R\right) + 2 \ln^2 \left(R\right) , \nonumber\\
S_1
&=&
15 + 3\zeta_2
  - \frac{2}{3}\zeta_3+ \left(14 + 2\zeta_2\right) 
\nonumber \\\nonumber
&& \times \ln \left(R\right)
+ 6 \ln^2 \left(R\right) +\frac{4}{3} \ln^3 \left(R\right), 
\end{eqnarray}
\begin{eqnarray}
\texttt{SE3l2M1md} &=&
(m^2)^{-2 \epsilon}
\Bigl[ 
\sum_{k=-2}^{1} S^d_k \epsilon^k + {\cal O} (\epsilon^2) 
\Bigr], 
\nonumber\\
 S_{-2}^d &=& \frac{1}{2},  \\
 S^d_{-1}&=& \frac{1}{2} \Bigl[1+ 2\ln \left(R\right)\Bigr], \nonumber \\
 S_0^d &=& \frac{1}{2}\left(1 + \zeta_2\right) +\ln \left(R\right) + \ln^2 \left(R\right),\nonumber \\
 S_1^d&=&
\frac{1}{6} \left(3 + 3\zeta_2 - 2\zeta_3\right)
+ \left(1 + \zeta_2\right)
\nonumber \\ \nonumber
&& \times \ln \left(R\right) 
+\ln^2 \left(R\right) + \frac{2}{3} \ln^3 \left(R\right).
\end{eqnarray}

For vertices, the external electrons are on their mass shell,
$p_i^2 = m^2$, $i=1,2$, and we introduce the Mandelstam invariant $s \equiv (p_1+p_2)^2$
(see Figure \ref{fig-nf2}).
Since each of the two vertices is related to two MIs,
we have to consider
\begin{eqnarray}\label{def:vert}
\texttt{V4l2M1m} &\equiv&\texttt{V4l2M1m}(1,1,1,1),\nonumber \\
\texttt{V4l2M1md}&\equiv&\texttt{V4l2M1m}(1,1,1,2),\nonumber\\
\texttt{V4l2M2m} &\equiv&\texttt{V4l2M2m}(1,1,1,1),\nonumber \\
\texttt{V4l2M2md}&\equiv&\texttt{V4l2M2m}(1,2,1,1).
\end{eqnarray}

We follow the same strategy employed for the sunrise diagrams.
First of all we derive the exact multi-dimensional MB representation,
and then we perform first a small-mass expansions in $m_s$, and then in $M_s$,
defined as the ratios of the fermion masses and the centre-of-mass
energy, $m_s \equiv -m^2 \slash s$,
$M_s \equiv -M^2 \slash s$.
The MB representations read as
\begin{multline}
\texttt{V4l2M1m} (\{\nu\}_4) =
(m^2)^{4-\nu_{1234}-2 \epsilon} 
\\
\times \frac{(-1)^{\nu_{1234}} e^{2 \gamma \epsilon}}{\prod_{i=1}^{4}
\Gamma({\nu_i})}
\int \frac{dz_1 dz_2}{(2\pi i)^2}
\\
\times m_s^{z_2-4+\nu_{1234}+2\epsilon}\,
 M_s^{-z_1+2-\nu_{12}-\epsilon}
\\
\times\frac{\prod_{i=1}^8 \Gamma_i}{\Gamma(2 z_1+\nu_{12})
\Gamma(z_1-\nu_{34}+4-2\epsilon)},\qquad
\end{multline}
with
\begin{xalignat}{2}
\Gamma_1&\equiv\Gamma(z_1+\nu_1),\nonumber\\
\Gamma_2&\equiv\Gamma(z_1+\nu_2),\nonumber\\
\Gamma_3&\equiv\Gamma(z_1+\nu_{12}-2+\epsilon),\nonumber\\
\Gamma_4&\equiv\Gamma(-z_2),\nonumber\\
\Gamma_5&\equiv\Gamma(2 z_2+\nu_4),\nonumber\\
\Gamma_6&\equiv\Gamma(-z_2-\nu_{34}+2-\epsilon),\nonumber \\
\Gamma_7&\equiv\Gamma(z_1-z_2-\nu_4+2-\epsilon),\nonumber\\
\Gamma_8&\equiv\Gamma(-z_1+z_2+\nu_{34}-2+\epsilon),
\end{xalignat}
and
\begin{multline}
\texttt{V4l2M2m}(\{\nu\}_4)=
(m^2)^{4-\nu_{1234}-2 \epsilon}
\\
\times \frac{ (-1)^{\nu_{1234}}e^{2 \gamma \epsilon}}
{\prod_{i=1}^4 \Gamma({\nu_i})}    \int \frac{dz_1 dz_2}{(2\pi i)^2}
\\
\times m_s^{z_2-4+\nu_{1234}+2\epsilon} \, M_s^{-z_1+2-\nu_{12}-\epsilon}
\\
\times \frac{\prod_{i=1}^9 \Gamma_i}
{\prod_{j=10}^{12} \Gamma_j},\qquad\qquad\qquad\qquad\qquad\qquad
\end{multline}
with
\begin{xalignat}{2}
\Gamma_1&\equiv\Gamma(-z_1),\nonumber\\
\Gamma_2&\equiv\Gamma(z_1+\nu_1),
\nonumber\\
\Gamma_3&\equiv\Gamma(z_1+\nu_2),\nonumber\\
\Gamma_4&\equiv\Gamma(z_1+\nu_{12}-2+\epsilon),
\nonumber\\
\Gamma_5&\equiv\Gamma(2 z_1-\nu_{34}+4-2 \epsilon),\nonumber\\
\Gamma_6&\equiv\Gamma(-z_2),
\nonumber\\
\Gamma_7&\equiv\Gamma(z_1-z_2-\nu_3+2-\epsilon),\nonumber\\
\Gamma_8&\equiv\Gamma(z_1-z_2-\nu_4+2-\epsilon),
\nonumber\\
\Gamma_9&\equiv\Gamma(-z_1+z_2+\nu_{34}-2+\epsilon),\nonumber\\
\Gamma_{10}&\equiv\Gamma(2 z_1+\nu_{12}),\nonumber\\
\Gamma_{11}&\equiv\Gamma(z_1-\nu_{34}+4-2\epsilon),\nonumber\\
\Gamma_{12}&\equiv \Gamma(2z_1-2z_2-\nu_{34}+4-2\epsilon).
\end{xalignat}

A careful analysis
of the powers of $m_s$ and $M_s$ under the MB integrals
leads to the following results,
\begin{eqnarray}
\texttt{V4l2M1m} &=&
(m^2)^{-2 \epsilon} 
\Bigl[ \sum_{k=-2}^0 V^1_{k} \epsilon^k + {\cal O}(\epsilon) \Bigr], \nonumber\\
V^{1}_{-2} &=& \frac{1}{2},\nonumber\\
V^{1}_{-1} &=& \frac{5}{2},\\
V^{1}_{0} &=& \frac{1}{2}\left[ 19 - 3\zeta_2 - \ln^2(m_s) \right],\nonumber
\end{eqnarray}
\begin{eqnarray}
\texttt{V4l2M1md} &=& 
 \frac{(m^2)^{-2 \epsilon}}{m^2}
\Bigl[ \sum_{k=-2}^0 V^{1d}_{k} \epsilon^k + {\cal O}(\epsilon) \Bigr], \nonumber\\
V^{1d}_{-2} &=& \frac{1}{2},\nonumber\\
V^{1d}_{-1} &=& 1+\frac{1}{2} \ln(m_s) ,\\
V^{1d}_{0} &=& 2-\zeta_2 +\ln(m_s)+\frac{1}{4}\ln^2(m_s),\nonumber
\end{eqnarray}
\begin{eqnarray}
\texttt{V4l2M2m} &=&
(m^2)^{-2 \epsilon} 
\Bigl[ \sum_{k=-2}^0 V^2_{k} \epsilon^k + {\cal O}(\epsilon) \Bigr], \nonumber\\
V^{2}_{-2} &=& \frac{1}{2},\nonumber\\
V^{2}_{-1} &=& \frac{5}{2} +\ln(m_s) ,\nonumber\\
V^{2}_{0} &=&  \frac{1}{2}(19+\zeta_2) +  5\ln(m_s)\nonumber\\
& +& \ln^2(m_s),
\end{eqnarray}
\begin{eqnarray}
\texttt{V4l2M2md} &=&
(m^2)^{-2 \epsilon} \frac{1}{s} 
\Bigl[ V^{2d}_{0} + {\cal O}(\epsilon) \Bigr], \nonumber\\
V^{2d}_{0} &=&
\frac{1}{6}\Bigl[ 12\zeta_3 - 6\zeta_2 \ln(M_s)\nonumber\\
 &-& \ln^3(M_s) \Bigr].
\end{eqnarray}

For box diagrams, the external momenta are again on their mass shell, 
and we have additionally $(p_1-p_3)^2 = t$.
After introducing $M_t \equiv - M^2 \slash t$, 
the appropriate MB representation is given by
\begin{multline}
\texttt{B5l2M2m}(\{\nu\}_5)=
(m^2)^{4-\nu_{12345}-2 \epsilon}
\\
\times \frac{(-1)^{\nu_{12345}}e^{2 \gamma \epsilon}}
{\prod_{i=1}^{5} \Gamma({\nu_i})}
 \int \frac{dz_1 dz_2 dz_3}{(2\pi i)^3}\\
\times M_t^{-z_1+2-\nu_{12}-\epsilon}
m_s^{z_3-4+\nu_{12345}+2 \epsilon}\\
\times \left(\frac{t}{s}\right)^{z_2-z_1+2-\nu_{12}-\epsilon}
 \frac{\prod_{i=1}^{11} \Gamma_i}{\prod_{j=12}^{14} \Gamma_j },\qquad
\end{multline}
with
\begin{xalignat}{2}
\Gamma_1&\equiv\Gamma(z_1+\nu_1),\\
\Gamma_2&\equiv\Gamma(z_1+\nu_2),\nonumber\\
\Gamma_3&\equiv\Gamma(z_1+\nu_{12}-2+\epsilon),\nonumber\\
\Gamma_4&\equiv\Gamma(-z_2),\nonumber\\
\Gamma_5&\equiv\Gamma(z_2+\nu_4), \nonumber \\
\Gamma_6&\equiv\Gamma(-z_3),\nonumber\\
\Gamma_7&\equiv\Gamma(-z_1+z_2),
\nonumber \\ \nonumber
\Gamma_8&\equiv\Gamma(2 z_1\!-\!2 z_2-\nu_{3445}+4-2 \epsilon),
\nonumber\\
\Gamma_{9}&\equiv\Gamma(z_1-z_2-z_3-\nu_{34}+2-\epsilon),\nonumber\\
\Gamma_{10}&\equiv\Gamma(z_1\!-\!z_2\!-\!z_3-\nu_{45}+2-\epsilon),
\nonumber\\
\Gamma_{11}&\equiv\Gamma(-z_1 +z_{2}+z_3+\nu_{345}-2+\epsilon),\nonumber\\
\Gamma_{12} &\equiv \Gamma(2 z_1+\nu_{12}),
\nonumber\\
\Gamma_{13} &\equiv \Gamma(z_1-\nu_{345}+4-2 \epsilon),\nonumber\\ \nonumber
\Gamma_{14} &\equiv \Gamma(2(z_1\!-\!z_2\!-\!z_3)-\nu_{3445}+4-2\epsilon).
\end{xalignat}
We have to compute two MIs,
\begin{eqnarray}\label{dfB}
\texttt{B5l2M2m}&\equiv&\texttt{B5l2M2m}(1,1,1,1,1),\nonumber \\
\texttt{B5l2M2md}&\equiv&\texttt{B5l2M2m}(1,2,1,1,1),
\end{eqnarray}
and an expansion in the high-energy limit of the appropriate
three-fold MB representations leads to the following results,
\begin{eqnarray}\label{def:box1}
\texttt{B5l2M2m} &=&
(m^2)^{-2 \epsilon} 
\Bigl[ \sum_{k=-2}^0 B_{k} \epsilon^k + {\cal O}(\epsilon) \Bigr], \nonumber\\
 B_{-2}&=& \frac{1}{s}\ln(m_s), \nonumber \\
 B_{-1}&=&
\frac{1}{ s}
 \Bigl(- \zeta_2 + 2 \ln(m_s) +  \frac{1}{2}\ln^2(m_s) \nonumber \\ &+& \ln(m_s)\ln(m_t)
\Bigr),
\nonumber
\end{eqnarray}
\begin{eqnarray}
 B_0&=&
%
%
%
\frac{1}{s}\Bigl[
-2 \zeta_2
- 2 \zeta_3
+ 4 \ln(m_s)
+ \ln^2(m_s)
\nonumber\\
&+& \frac{1}{3} \ln^3(m_s)
- 4 \zeta_2 \ln(m_t)\nonumber\\
&+& 2 \ln(m_s)\ln(m_t)
+ \ln(m_s)\ln^2(m_t)
\nonumber\\
&-&\frac{1}{6}\ln^3(m_t)\nonumber\\
&-& \Bigl( 3 \zeta_2 
+ \frac{1}{2} \ln^2(m_s)
- \ln(m_s) \ln(m_t)
 \nonumber\\
&+& \frac{1}{2} \ln^2(m_t) \Bigr) \ln\left( 1+ \frac{t}{s}\right)\nonumber\\
&-& \Bigl( \ln(m_s) -\ln(m_t) \Bigr) \text{Li}_2 \left(-\frac{t}{s}\right)
\nonumber\\
&+& \text{Li}_3 \left(-\frac{t}{s}\right)
\Bigr],
\end{eqnarray}
%
\begin{eqnarray}\label{def:box2}
\texttt{B5l2M2md} &=&
(m^2)^{-2 \epsilon} 
\Bigl[ \sum_{k=-1}^0 B^d_{k} \epsilon^k + {\cal O}(\epsilon) \Bigr], 
\nonumber\end{eqnarray}
\begin{eqnarray}
 B^d_{-1}&=&
       -\frac{1}{st} \Bigl[\ln(m_s)\ln(m_t)-\ln(m_s)L(R)\Bigr],
 \nonumber\\
 B^d_0&=&
 \frac{1}{st}\Bigl\{
-2 \zeta_3
+ \zeta_2 \ln(m_s)
+ 4 \zeta_2 \ln(m_t)\nonumber\\
&-& 2 \ln(m_s)\ln^2(m_t)
+ \frac{1}{6}\ln^3(m_t)
   \nonumber  \\
&-& 2 \zeta_2 L(R)
+ 2 \ln(m_s) \ln(m_t)L(R)
\nonumber\\
&-& \frac{1}{6}L^3(R)
\\
&+& \Bigl( 3 \zeta_2  
+ \frac{1}{2} \ln^2(m_s) -  \ln(m_s) \ln(m_t)
\nonumber\\
&+& \frac{1}{2}\ln^2(m_t)  \Bigr) \ln\left( 1+ \frac{t}{s}\right)\nonumber\\
&+& \Bigl( \ln(m_s) -\ln(m_t) \Bigr) \text{Li}_2 \left(-\frac{t}{s}\right)
\nonumber\\
&-& \text{Li}_3 \left(-\frac{t}{s}\right)
\Bigr\}.\nonumber
\end{eqnarray}

\section{Summation techniques\label{section-techn}}

In any realistic computation we have to check the
structure of the ultraviolet (UV) and infrared (IR) divergencies.
By combining the Mellin-Barnes method with recently developed
summation techniques we are able to evaluate exactly 
(i.e. without a high-energy approximation) the residues of the UV and IR poles
for each MI.

A simple example is enough to illustrate our procedure. 
We consider
the following one-fold integral in the complex plane,
related to the single pole of \texttt{V4l2M1md},
\begin{equation}
I
\equiv
\frac{1}{2\pi i} \int_{c-i\infty}^{c+i \infty} d z \ M_s^{-z}\
\frac{ \prod_{i=1}^3 \Gamma_i }{\Gamma(2 z + 2)},
\end{equation}
where we recall that $M_s \equiv - M^2 \slash s $, the integration contour is a straight
line parallel to the imaginary axis, $c= - 1 \slash 2$
and we introduced
\begin{xalignat}{2}
\Gamma_1&\equiv \Gamma(-z),\nonumber\\
\Gamma_2&\equiv \Gamma(z),\nonumber\\
\Gamma_3&\equiv \Gamma^2(z+1).
\end{xalignat}
After closing the integration contour to the right of the complex plane and taking
residua, the integral $I$ can be written by means of two inverse binomial sums,
\begin{eqnarray}
I&=&
\sum_{n=1}^{\infty}
\left(-1 \right)^n\ M_s^{-n}
\frac{1}{\binom{2 n}{n}} \left( \frac{1}{n} - \frac{2}{2 n+1}\right)\nonumber\\
&-& 2 - \ln\left(M_s \right).
\end{eqnarray}
Inverse binomial sums were recently studied by means of the 
log-sine approach in \cite{Davydychev:2003mv}.
Another approach was developed in \cite{Weinzierl:2004bn}
by generalizing the summation algorithms
introduced in \cite{Moch:2001zr}.
A straightforward application of these techniques leads to
a compact result,
\begin{equation}
I
=
\frac{1-x_M}{1+x_M} \ln (y_M) \left(1+4 M_s \right)
-
\ln \left(M_s \right),
\end{equation}
where we introduced the variables $x_M$ and $y_M$,
\begin{eqnarray}
x_M &\equiv& \frac{\sqrt{4 M^2 -s} - \sqrt{-s}}{\sqrt{4 M^2 -s} + \sqrt{-s}},\nonumber\\
y_M &\equiv& \frac{\sqrt{4 M^2 -t} - \sqrt{-t}}{\sqrt{4 M^2 -t} + \sqrt{-t}}.
\end{eqnarray}

As an example, after additionally introducing the following variables,
\begin{eqnarray}
x_m &\equiv& \frac{\sqrt{4 m^2 -s} - \sqrt{-s}}{\sqrt{4 m^2 -s} + \sqrt{-s}},\nonumber\\
y_m &\equiv& \frac{\sqrt{4 m^2 -t} - \sqrt{-t}}{\sqrt{4 m^2 -t} + \sqrt{-t}},
\end{eqnarray}
we get the non-approximated
expressions for the residues of the poles of the two box diagrams
defined in Eq.~\eqref{dfB},
\begin{eqnarray}
B_{-2}&=& - \frac{1}{m^2} \frac{x_m}{1-x_m^2} H(0;x_m),\nonumber \\
B_{-1}&=& 
\frac{1}{2 m^2} \frac{x_m}{1-x_m^2}
\Bigl\{-
H^2(0;x_m) \nonumber\\
&+& 2 \Bigl[
\zeta_2 - 2 H(0,-1;x_m)
\Bigr]
\nonumber\\
&+& 2 H(0;x_m)
\Bigl[
2 H(-1;x_m)\nonumber\\
&-&\frac{1+y_M}{1-y_M} H(0;y_M)
-
2 \nonumber \\
&-& \ln \left(\frac{m^2}{M^2}
\right)
\Bigr]
\Bigr\},
\end{eqnarray}
and 
\begin{eqnarray}
 B^d_{-1}&=& - \frac{1}{m^2 M^2} \frac{x_m\,y_M}{(1-x_m^2)(1-y_M^2)}\nonumber \\
 &\times&H(0;x_m) H(0;y_M),
\end{eqnarray}
where we used the HPLs introduced in \cite{Remiddi:1999ew}.

For completeness, we add here also the exact expressions for the diveregent parts of the vertex MIs:
\begin{eqnarray}
  V^{1d}_{-1} &=&  \frac{1}{2}
  \left\{
  \frac{1+x_M}{1-x_M} H(0;x_M) +2 + \ln R
  \right\}
\nonumber \\
 V^{2}_{-1} &=&  \frac{5}{2} + \frac{1+x_m}{1-x_m} H(0;x_m).
\end{eqnarray}

\section{Summary}
From \cite{Czakon:2004wm} we know the table of MIs for massive
two-loop Bhabha scattering. 
We were able to express all
the Feynman integrals occurring in the amplitude
through these MIs by algebraic relations.  
We presented at the workshop all the planar two-loop box MIs.
The $N_f=1$ MIs has been published in the meantime \cite{Czakon:2006pa}.
In this contribution, we provide the expanded results for all the MIs
entering the Bhabha-scattering amplitude with two fermion flavors in
the limit of small fermion masses at fixed scattering angle.
The MIs may be also found at our webpage \cite{web-masters:2006nn}.
These MIs were one of the last missing ingredients for the evaluation of the
virtual two-loop contribution to the differential cross-section.  

The computation of the last nine
non-planar two-loop box MIs is under way.

\section*{Acknowledgements}
We would like to thank S. Moch for useful discussions.

Work supported in part by Sonderforschungsbereich/Transregio 9--03 of DFG
`Computergest{\"u}tzte Theo\-re\-ti\-sche Teil\-chen\-phy\-sik',  by
the Sofja Kovalevskaja Award of the Alexander von Humboldt Foundation
  sponsored by the German Federal Ministry of Education and Research,
and by the Polish State Committee for Scientific Research (KBN),
research projects in 2004--2005.

\end{document}